\pdfoutput=1

\def\arxiv{1}

\if\arxiv1
\PassOptionsToClass{nonacm}{acmart}
\fi

\documentclass[sigplan,screen]{acmart}

\setcopyright{acmlicensed}
\acmPrice{15.00}
\acmDOI{10.1145/3385412.3386024}
\acmYear{2020}
\copyrightyear{2020}
\acmSubmissionID{pldi20main-p713-p}
\acmISBN{978-1-4503-7613-6/20/06}
\acmConference[PLDI '20]{Proceedings of the 41st ACM SIGPLAN International Conference on Programming Language Design and Implementation}{June 15--20, 2020}{London, UK}
\acmBooktitle{Proceedings of the 41st ACM SIGPLAN International Conference on Programming Language Design and Implementation (PLDI '20), June 15--20, 2020, London, UK}

\usepackage[utf8]{inputenc}
\usepackage[T1]{fontenc}
\usepackage[english]{babel}

\usepackage{tikz}                            % For drawings
\usepackage[nomain,acronyms]{glossaries}     % For automated acronyms
\usepackage{booktabs}                        % For formal tables
\usepackage{subcaption}                      % For complex figures with subfigures/subcaptions
\usepackage[printwatermark]{xwatermark}      % For version watermark
\usepackage{blindtext}                       % For blind text
\usepackage[capitalise, noabbrev]{cleveref}  % For nice references
\usepackage{tabularx}                        % For fixed-width tables
\usepackage{listings}                        % For source code listings
\usepackage[binary-units]{siunitx}           % For nice SI units
\usepackage{microtype}                       % For the magic touch
\usepackage{xifthen}                         % For Alan
\usepackage{graphicx}                        % For graphics
\usepackage{threeparttable}                  % For nice footnotes in tables
\usepackage{microtype}                       % For improved typesetting
\graphicspath{{figures/}}

\definecolor{keyword}{rgb}{0,0.5,0.75}
\definecolor{revision}{rgb}{0.0, 0.4, 0.8}

\def\revcolor{revision}

\lstdefinelanguage{llhd}{
    morekeywords = {func,proc,entity,const,alias,mux,reg,insf,inss,extf,exts,del,call,con,inst,drv,st,halt,ret,br,wait,not,neg,sig,prb,var,ld,add,sub,and,or,xor,smul,sdiv,smod,srem,umul,udiv,umod,urem,eq,neq,slt,sgt,sle,sge,ult,ugt,ule,uge,shl,shr,phi,for,low,high,rise,fall,both,declare,with,alloc,free},
    morecomment = [l]{;},
}
\floatstyle{plaintop}
\newfloat{listing}{tbp}{lst}
\floatname{listing}{Listing}

\crefname{section}{\S}{\S\S}
\Crefname{section}{\S}{\S\S}

\newcommand{\ifequals}[3]{\ifthenelse{\equal{#1}{#2}}{#3}{}}
\newcommand{\llhdlower}{llhd}
\newcommand{\llhd}[1][]{%
  \ifthenelse{\isempty{#1}}{%
    LLHD%
  }{%
    \ifequals{#1}{1}{Behavioural}%
    \ifequals{#1}{2}{Structural}%
    \ifequals{#1}{3}{Netlist}
    \llhd{}%
  }%
}
\newcommand{\moore}{Moore} % :-)
\newcommand{\llhdsim}{\llhd{}-Sim}
\newcommand{\llhdblaze}{\llhd{}-Blaze}

\newcommand*\rot{\rotatebox{90}}

\newcolumntype{L}{>{\raggedright\arraybackslash}X}
\newcolumntype{C}{>{\centering\arraybackslash}X}
\newcolumntype{R}{>{\raggedleft\arraybackslash}X}

\newcommand{\secref}[1]{{Section~\ref{#1}}}

\usetikzlibrary{shapes}
\usetikzlibrary{arrows}
\usetikzlibrary{positioning}

\glsdisablehyper
\newacronym{ast}{AST}{Abstract Syntax Tree}
\newacronym{cdc}{CDC}{Clock Domain Crossing}
\newacronym{cfg}{CFG}{Control Flow Graph}
\newacronym{cf}{CF}{Constant Folding}
\newacronym{cgra}{CGRA}{Coarse-Grained Reconfigurable Array}
\newacronym{cmos}{CMOS}{Complementary Metal–Oxide–Semiconductor}
\newacronym{cse}{CSE}{Common Subexpression Elimination}
\newacronym{ctl}{CTL*}{Computational Tree Logic}
\newacronym{dce}{DCE}{Dead Code Elimination}
\newacronym{dfg}{DFG}{Data Flow Graph}
\newacronym{dnf}{DNF}{Disjunctive Normal Form}
\newacronym{eda}{EDA}{Electronic Design Automation}
\newacronym{fifo}{FIFO}{First-In First-Out}
\newacronym{fir}{FIR}{Finite Impulse Response}
\newacronym{fpga}{FPGA}{Field-Programmable Gate Array}
\newacronym{gcse}{GCSE}{Global Common Subexpression Elimination}
\newacronym{hdl}{HDL}{Hardware Description Language}
\newacronym{hls}{HLS}{High-level Synthesis}
\newacronym{ir}{IR}{Intermediate Representation}
\newacronym{is}{IS}{Instruction Simplification}
\newacronym{jit}{JIT}{just-in-time}
\newacronym{lec}{LEC}{Logical Equivalence Checker}
\newacronym{ecm}{ECM}{Early Code Motion}
\newacronym{licm}{LICM}{Loop-Invariant Code Motion}
\newacronym{ltl}{LTL}{Linear Temporal Logic}
\newacronym{mlir}{MLIR}{Multi-Level Intermediate Representation}
\newacronym{pl}{PL}{Process Lowering}
\newacronym{soc}{SoC}{System on Chip}
\newacronym{ssa}{SSA}{Static Single Assignment}
\newacronym{tcfe}{TCFE}{Total Control Flow Elimination}
\newacronym{tcm}{TCM}{Temporal Code Motion}
\newacronym{trg}{TRG}{Temporal Region Graph}
\newacronym{tr}{TR}{Temporal Region}

\begin{document}

\title{LLHD: A Multi-level Intermediate Representation for Hardware Description Languages}

%%%%%%%%%%%%%%%%%
%%   Authors   %%
%%%%%%%%%%%%%%%%%

\author{Fabian Schuiki}
\orcid{0000-0002-9923-5031}
\affiliation{
  \department{Integrated Systems Laboratory (IIS)}
  \institution{ETH Zürich}
  \streetaddress{Gloriastrasse 35}
  \city{Zürich}
  \postcode{8092}
  \country{Switzerland}
}
\email{fschuiki@iis.ee.ethz.ch}

\author{Andreas Kurth}
\orcid{0000-0001-5613-9544}
\affiliation{
  \department{Integrated Systems Laboratory (IIS)}
  \institution{ETH Zürich}
  \streetaddress{Gloriastrasse 35}
  \postcode{8092}
  \city{Zürich}
  \country{Switzerland}
}
\email{akurth@iis.ee.ethz.ch}

\author{Tobias Grosser}
\affiliation{
  \department{Scalable Parallel Computing Laboratory (SPCL)}
  \institution{ETH Zürich}
  \streetaddress{Universitätstrasse 6}
  \postcode{8092}
  \city{Zürich}
  \country{Switzerland}
}
\email{tobias.grosser@inf.ethz.ch}

\author{Luca Benini}
\orcid{0000-0001-8068-3806}
\affiliation{
  \department{Integrated Systems Laboratory (IIS)}
  \institution{ETH Zürich}
  \streetaddress{Gloriastrasse 35}
  \postcode{8092}
  \city{Zürich}
  \country{Switzerland}
}
\email{lbenini@iis.ee.ethz.ch}

\begin{abstract}
Modern \glspl{hdl} such as SystemVerilog or VHDL are, due to their sheer complexity, insufficient to transport designs through modern circuit design flows.
Instead, each design automation tool lowers \glspl{hdl} to its own \gls{ir}.
These tools are monolithic and mostly proprietary, disagree in their implementation of \glspl{hdl}, and while many redundant \glspl{ir} exists, no \gls{ir} today can be used through the entire circuit design flow.
To solve this problem, we propose the \llhd{} multi-level \gls{ir}.
\llhd{} is designed as simple, unambiguous reference description of a digital circuit, yet fully captures existing \glspl{hdl}.
We show this with our reference compiler on designs as complex as full CPU cores.
\llhd{} comes with lowering passes to a hardware-near structural \gls{ir}, which readily integrates with existing tools.
\llhd{} establishes the basis for innovation in \glspl{hdl} and tools without redundant compilers or disjoint \glspl{ir}.
For instance, we implement an \llhd{} simulator that runs up to $2.4\times$ faster than commercial simulators but produces equivalent, cycle-accurate results.
An initial vertically-integrated research prototype is capable of representing all levels of the IR, implements lowering from the behavioural to the structural IR, and covers a sufficient subset of SystemVerilog to support a full CPU design.

\end{abstract}

%%%%%%%%%%%%%%%%%%
%%   Keywords   %%
%%%%%%%%%%%%%%%%%%

%% 2012 ACM Computing Classification System (CSS) concepts
%% Generate at 'http://dl.acm.org/ccs/ccs.cfm'.

%% End of generated code

%% Keywords
%% comma separated list

%%%%%%%%%%%%%%%%%
%%   Content   %%
%%%%%%%%%%%%%%%%%

\maketitle
\section{Introduction}
\label{sec:intro}

% Thanks Simon Peyton Jones for being the most awesome paper-writing functional-programming-guru guy ever!

\begin{figure}
    \includegraphics[width=\linewidth]{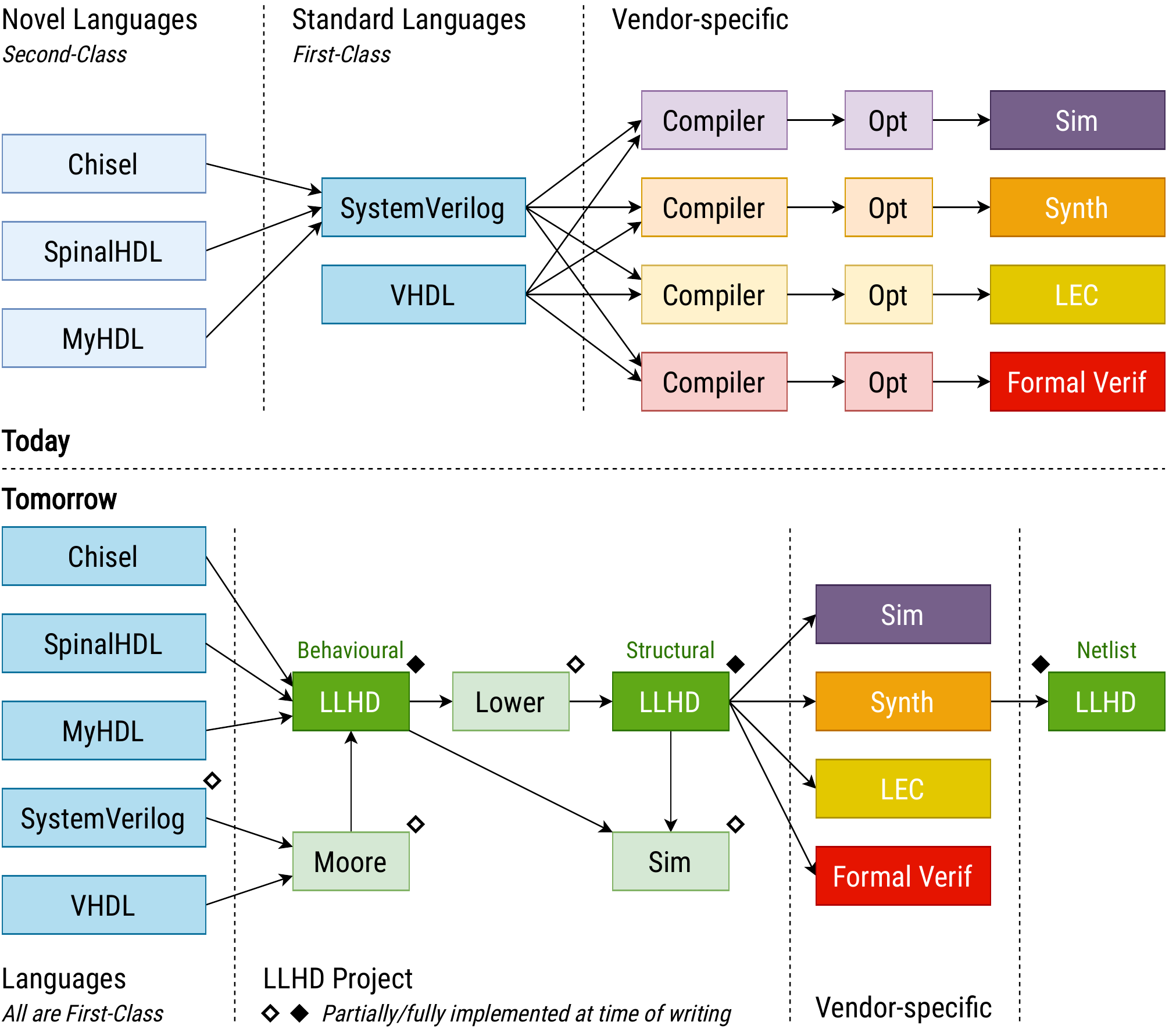}
    \caption{
        Redundancy in today's hardware design flow (top).
        Replacement flow with \moore{} as compiler frontend and \llhd{} as unifying \acrshort{ir} (bottom).
        Maturity of the implementation at the time of writing is indicated.
    }
    \label{fig:intro}
\end{figure}

% The problem.
The workflow we use today to design digital circuits, including CPUs and GPUs, has a severe redundancy problem.
The sheer complexity of modern designs, requiring billions of transistors to be placed and costing millions of dollars on each silicon iteration, has given rise to a very deep design toolchain --- much more so than in software development.
Along their way to becoming silicon, digital designs pass simulators, formal verifiers, linters, synthesizers, logical equivalence checkers, and many other tools.
\Glspl{hdl} are used as the input to these tools, with SystemVerilog and VHDL being widely used in industry.
Both are highly complex languages: their specifications span 1275 and 626 pages of text, respectively~\cite{sv2017lrm, vhdl2008lrm}.
The aforementioned tools come from different vendors (often \emph{by design} to rule out systematic sources of error), each of which has its own implementation of the complex language standards, as depicted in \cref{fig:intro}.
Hardware designers have to ``hope'', for example, that the circuit synthesizer interprets the semantics of a design in the same way as the simulator used for verification, and the standards can be very subtle~\cite{sutherland2006synthesizable}.
As a consequence, designers resort to a ``safe subset'' of each language known to produce identical results across most tools, which precludes the practical use of many distinguishing high-level features of SystemVerilog and VHDL.

% It's an interesting problem.
All of our modern computing infrastructure depends on getting this tool flow right.
As a consequence, the \gls{eda} industry invests significant resources into finding safe subsets of the languages, establishing coding styles, and developing linting tools to enforce adherence to such conventions~\cite{sutherland2006synthesizable}.
In contrast to software development, where the advent of \glspl{ir} and frameworks such as LLVM~\cite{lattner2004llvm} has provided a productive platform for open development, the hardware design flow remains isolated and vendor-locked.
The \gls{eda} market is dominated by closed-source, proprietary, monolithic toolchains, which have been highly optimized over decades.
New open-source tools face the hurdle of implementing complex language frontends before being able to compete.
We observe that hardware engineering and compiler design communities have evolved in isolation, even though many optimizations and methodological improvements readily apply to both worlds.
This is to hardware engineering's severe disadvantage, as we see ample opportunity for the two communities to connect, exchange, and benefit from each other.

% Here is my idea.
We propose \emph{\llhd{}}, an \gls{ir} to represent digital circuits throughout the \emph{entire} design flow, from simulation, testbenches and formal verification, behavioural and structural modeling, to synthesis and the final gate-level netlist.
There cannot be a single IR that fits all hardware needs, as this would require high-level simulation constructs without hardware equivalents, yet still be trivially synthesizable.
However, the constructs needed to describe a netlist form a subset of those needed for synthesis, which are in turn a subset of those needed for simulation.
As such, \llhd{} is a multi-level \gls{ir} with three distinct levels or dialects, which cater to the corresponding parts of the tool flow.
\llhd{} adheres to \gls{ssa} form~\cite{alpern1988detecting,cytron1989efficient}, which lends itself exceptionally well to represent digital circuits, which are exactly that: signals with single, static driver assignments.
\llhd{} borrows the basic \gls{ir} syntax from LLVM but defines an \gls{ir} for digital circuits, which must explicitly deal with the passing of time and, since digital circuits are inherently concurrent, must be able to describe concurrency.
Together with \emph{\moore{}}, a compiler frontend for \glspl{hdl}, \llhd{} significantly reduces redundancy in the design flow and allows novel, open languages to thrive, as depicted in \cref{fig:intro}.

% It's an unsolved problem.
Many \glspl{ir} for hardware designs already exist.
For instance, \gls{eda} tools have internal \glspl{ir}, but these are highly tool-specific and mostly proprietary.
In general, the vast majority of \glspl{ir} puts a narrow focus on circuit synthesis.
To prevent redundancy and disagreement in compilers and gaps between \glspl{ir}, we argue that one \gls{ir} must cover the entire circuit design flow.
To our knowledge this paper is the first to propose this solution.
We make the following contributions:

% Contributions
\begin{itemize}
    \item We define a multi-level \gls{ir} that captures current \glspl{hdl} in an \gls{ssa}-based form compatible with modern, imperative compilers but with extensions and specializations crucial to represent digital hardware~(\cref{sec:lang}).
    \item We show how existing industry-standard \glspl{hdl}, such as SystemVerilog and VHDL, map to this \gls{ir}~(\cref{sec:map}).
    \item We establish transformation passes to lower from \llhd[1]{} to hardware-near \llhd[2]{}~(\cref{sec:opt}).
    \item We show that such a multi-level \gls{ir} can improve the existing \gls{eda} tool flow, even without explicit support by commercial tools~(\cref{sec:flow}).
    \item We provide evidence that the \gls{ir} can capture complex designs such as entire CPU cores \cite{zaruba2020snitch}, that a minimal reference simulator models those designs identically to commercial simulators, and that an early optimized simulator runs up to $2.4 \times$ faster than a commercial simulator~(\cref{sec:eval}).
\end{itemize}

Finally, we provide an open-source implementation of our \gls{ir}, its reference simulator, and an accompanying \gls{hdl} compiler.\footnote{Project Website: \url{http://llhd.io/}}
The implementation acts as a vertically-integrated research prototype.
This prototype is currently capable of capturing behavioural, structural, and netlist \llhd{}.
Lowering from behavioural to structural \llhd{} is partially implemented in order to demonstrate the key transformations, but is not complete at the time of writing.
Lowering from structural to netlist \llhd{} is the domain of hardware synthesizers and as such outside the scope of this work.
The \moore{} compiler supports a subset of SystemVerilog which is large enough to represent a full CPU core \cite{zaruba2020snitch} and covers a sufficient amount of non-synthesizable constructs of the language to support simple testbenches.
Initial experimental work on VHDL support is underway.
Our simulator implementation covers the vast majority of all three \llhd{} dialects, except a few instruction that were not instrumental to simulate the designs presented in this work.

\section{The \llhd{} \acrlong{ir}}
\label{sec:lang}

The \llhd{} \gls{ir} is designed as an \gls{ssa} language~\cite{alpern1988detecting,cytron1989efficient} which enables a very direct representation of the data flow in a digital circuit.
The rationale is that modern digital circuits are essentially the same as an \gls{ssa} data flow graph, where each logic gate corresponds to a node in the graph.
\llhd{} has an in-memory representation, a human-readable representation, and a planned binary on-disk representation, and all three representations are equivalent.
In this section, we define the core concepts of the \llhd{} language; the complete and precise specification is part of the \llhd{} Language Reference Manual.\footnote{Language Reference Manual: \url{http://llhd.io/spec.html}}

%-------------------------------------------------------------------------------
\subsection{Describing Digital Circuits}
\label{sec:lang_essence}

\begin{figure}
    \centering
    \includegraphics[width=\linewidth]{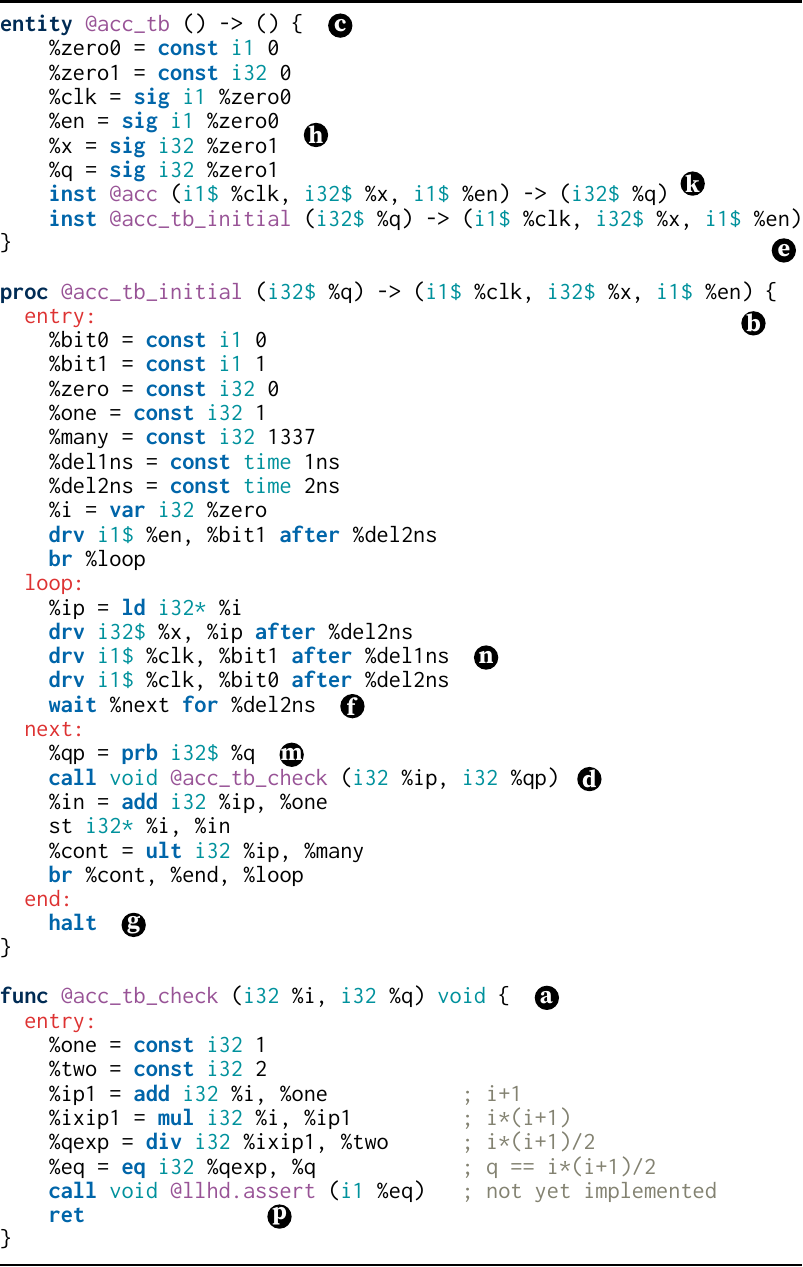}
    \caption{
        A testbench for an accumulator design as an illustrative example for \llhd{} code.
        See \cref{sec:lang} for a detailed description, \cref{fig:map} for the corresponding SystemVerilog code, and \cref{fig:opt_example} for the implementation of \texttt{@acc}.
    }
    \label{fig:lang}
\end{figure}

A hardware description requires a notion of passing time, must be able to represent concurrency, and provide a way to describe the structure and hierarchy of a circuit.
This is due to digital circuits being inherently hierarchical, concurrent, and time-dependent.
\llhd{} draws from the abstractions established in modern \glspl{hdl} over the past decades, and distills them into fundamental orthogonal concepts.
\Cref{fig:lang} shows a sample \llhd{} source text describing a testbench for an accumulator circuit as an illustrative example.
The language provides three modeling constructs:
\begin{description}
    \item[Functions] capture a mapping from a set of input values to an output and allow for reuse of computation.
    They facilitate code reuse, recursion and program-defined mapping of a set of input values to a singular output value in the \gls{ssa} graph, but they do not have a direct hardware equivalent.
    (See \cref{fig:lang}a)
    \item[Processes] are Turing-complete subprograms that describe how a circuit's state and output reacts to a change in its input.
    This provides a \emph{behavioural} circuit description.
    (See \cref{fig:lang}b)
    \item[Entities] build hierarchy by instantiating other processes or entities, which then operate concurrently.
    This provides a \emph{structural} circuit description.
    Such instantiation translates into reuse by replication in the physical silicon, which is an essential contributor to our ability to manufacture designs with billions of transistors.
    (See \cref{fig:lang}c)
\end{description}
Circuit designs written in \glspl{hdl} generally model circuits behaviourally and structurally.
Physical silicon itself is a purely structural arrangement of circuits and thus fully captured by a hierarchy of entities.
Functions and processes are merely modeling tools to fully capture \glspl{hdl}, including simulation and verification components which have no physical equivalent.
To capture these levels of abstraction, \llhd{} is a three-level \gls{ir}, as described in the following paragraph.

%-------------------------------------------------------------------------------
\subsection{Multi-level Intermediate Representation}
\label{sec:lang_mlir}

\llhd{} can capture \emph{all} aspects of a digital design written in a higher-level \gls{hdl} such as SystemVerilog or VHDL, including simulation, verification, and testing constructs.
However, its structure also allows it to clearly capture parts relevant for synthesis, as well as represent the netlist that results from synthesis.
This makes \llhd{} a \emph{multi-level \gls{ir}} with the following levels:
\begin{description}
    \item[{\llhd[1]{}}] aims at capturing circuit descriptions in higher-level \glspl{hdl} as easily as possible.
    It allows for simulation constructs and test benches to be fully represented, including assertions, file I/O, or formal verification information as intrinsics.

    \item[{\llhd[2]{}}] limits the description to the parts that describe the input to output relations of a design.
    This covers essentially everything that can be represented by an entity (see \cref{sec:opt} for a more technical description).

    \item[{\llhd[3]{}}] further limits the description to just entities and instructions to instantiate and connect sub-circuits.
    More specifically allowed are just the \texttt{entity} construct, as well as signal creation (\texttt{sig}), connection (\texttt{con}), delay (\texttt{del}), and sub-circuit instantiation (\texttt{inst}).
\end{description}
We observe that the constructs of \llhd[3]{} are a strict subset of \llhd[2]{}, which in turn is a strict subset of \llhd[1]{}.
Rather than defining three separate \glspl{ir}, we thus propose one holistic \gls{ir} to cover the entire process.
As a design makes its way through the hardware design flow, the simulation and design verification phase uses the full \gls{ir} of \llhd[1]{}.
Before synthesis, \llhd{} compiler passes lower the design to \llhd[2]{}.
A synthesizer then lowers the design to a netlist by performing logic synthesis and mapping the design to a target silicon technology, which can be expressed in \llhd[3]{}.

The remainder of this section provides a more detailed treatment of the constructs in \llhd{}.
In the following two sections, we first discuss mapping a design from a \gls{hdl} to \llhd{} (\cref{sec:map}) and then the compiler passes to transform the design between the different \llhd{} levels (\cref{sec:opt}).

%-------------------------------------------------------------------------------
\subsection{Modules, Names, and Types}
\label{sec:lang_mod}

A single \llhd{} source text is called a \emph{module}.
Modules consist of functions, processes, and entities.
Multiple modules can be combined by a linker, which resolves references in one module against the definitions made in the other.

\llhd{} distinguishes between three types of names.
Most importantly, to minimize naming conflicts in complex designs, only \emph{global names}, such as \verb|@foo|, are visible to other modules during linking.
\emph{Local names}, such as \verb|%bar|, and anonymous names, such as \verb|%42|, are only visible within the current module (for functions, processes, and entities) or the current unit (for values).

\llhd{} is a strongly-typed language, i.e., every value must have a type.
\llhd{} supports a set of types typical for an imperative compiler: \verb|void| (no value), \verb|iN| (\verb|N|-bit integers), \verb|T*| (pointer to type \verb|T|), \verb|[N x T]| (array of \verb|N| elements of type \verb|T|, and \verb|{T1,T2,...}| (structure with fields of type \verb|T1| etc.).
\llhd{} defines the following hardware-specific types:
\begin{description}
    \item[\texttt{time}] represents a point in time.
    This allows to describe delays (e.g., through gates) and elapsed time (e.g., in simulation).
    \item[\texttt{nN}] is an enumeration value that can take one of \texttt{N} distinct values.
    This allows to represent non-power-of-two values (e.g., the set of states in a state machine or the inputs to a multiplexer).
    \item[\texttt{lN}] is a nine-valued logic value, defined in the IEEE~1164 standard~\cite{ieee1164}.
    This allows to model states that a physical signal wire may be in (drive strength, drive collision, floating gates, and unknown values).
    % The \texttt{ld} and \texttt{st} instructions are used to read and modify the value at the pointed-to location.
    \item[\texttt{T\$}] is a signal carrying a value of type \texttt{T}.
    This represents a physical signal wire.
    The \texttt{prb} and \texttt{drv} instructions are used to read the current value of the signal and trigger a future change of the signal.
\end{description}

%-------------------------------------------------------------------------------
\subsection{Units}
\label{sec:lang_unit}

\begin{table}
    \caption{Overview of the design units available in \llhd{} with their execution paradigm and timing model. See \cref{sec:lang_essence} and \cref{sec:lang_unit} for a detailed description.}
    \small
    \begin{tabularx}{\linewidth}{@{}lllL@{}}
        \toprule
        \textbf{Unit} & \textbf{Execution} & \textbf{Timing} & \textbf{Use} \\
        \midrule
        Function & control flow & immediate & user-def. \gls{ssa} mapping \\
        Process  & control flow & timed     & behavioural circ. desc. \\
        Entity   & data flow    & timed     & structural circ. desc. \\
        \bottomrule
    \end{tabularx}
    \label{tab:lang_unit}
\end{table}

The three main constructs of \llhd{}, functions, processes, and entities, are called \emph{units}.
As shown in \cref{tab:lang_unit}, \llhd{} defines for each unit how instructions are executed (\emph{execution paradigm}) and how time passes during the execution of the unit (\emph{timing model}).

The execution paradigm is either control or data flow.
\emph{Control flow} units consist of basic blocks, where execution follows a clear control flow path.
Each basic block must have exactly one terminator instruction which transfers control to another basic block, to the caller, or halts completely.
\emph{Data flow} units consist only of a set of instructions which form a \gls{dfg}.
Execution of instructions is implied by the propagation of value changes through the graph.

The timing model can be immediate or timed.
\emph{Immediate} units execute in zero time.
They may not contain any instructions that suspend execution or manipulate signals.
These units are ephemeral in the sense that their execution starts and terminates in between physical time steps.
As such no immediate units coexist or persist across time steps.
\emph{Timed} units coexist and persist during the entire execution of the \gls{ir}.
They represent reactions to changes in signals and as such model the behaviour of a digital circuit.
Such units may suspend execution or interact with signals by probing their value or scheduling state changes.

\subsubsection{Functions}
\label{sec:lang_func}

Functions represent a mapping from zero or more input arguments to zero or one return value.
Functions are defined with the \texttt{func} keyword and are called from other units with the \texttt{call} instruction.
For example, \cref{fig:lang}a defines a function to assert that an input value \texttt{\%q} matches the sum of all integers up to \texttt{\%i}, and \cref{fig:lang}d calls that function.
Functions execute immediately, meaning that they cannot interact with signals or suspend execution.

\subsubsection{Processes}
\label{sec:lang_proc}

Processes, in contrast, can interact with time.
Similar to functions, they are executed in a control flow manner.
Processes represent a circuit with zero or more inputs and outputs.
The inputs and outputs must be of a signal type \texttt{T\$}; other types are not permitted.
\cref{fig:lang}b defines a process that generates the patterns to test an accumulator circuit, and \cref{fig:lang}e instantiates that process.
Upon initialization, control starts at the first basic block and proceeds as it would in a function.
Processes may probe and drive the value of signals (see \cref{sec:lang_isa_sig}), which are a process' only means to communicate with other parts of the design.
Additionally, processes may suspend execution for a period of time or until a signal change (\texttt{wait}, \cref{fig:lang}f), or indefinitely (\texttt{halt}, \cref{fig:lang}g).
In contrast to functions, processes exist throughout the entire lifetime of the circuit and never return.

\subsubsection{Entities}
\label{sec:lang_entity}

Entities describe a pure \gls{dfg} and their execution is not governed by any control flow.
Upon initialization, all instructions are executed once.
At all subsequent points in time, instructions are re-executed if one of their inputs changes.
This creates an implicit execution schedule for the instructions.
Entities build structure and design hierarchies by allocating registers and signals via the \texttt{reg} and \texttt{sig} instructions, and instantiating other entities and processes via the \texttt{inst} instruction.
For example, \cref{fig:lang}c defines an entity which has four local signals (h), and instantiates the \texttt{@acc} design to be tested (k) and the \texttt{@acc\_tb\_initial} process to execute the test (e).

%-------------------------------------------------------------------------------
\subsection{Instruction Set}
\label{sec:lang_isa}

\llhd{}'s simple instruction set captures the essence of hardware descriptions at a hardware-near level of abstraction.
Nevertheless, \llhd{} preserves arithmetic operations, which are the main target of many optimizations in commercial hardware synthesizers.
As a general rule, all instructions contain sufficient type annotations to determine the type of each operand and the result.
We omit a detailed description of all instructions, especially those that are common in imperative compiler \glspl{ir} such as LLVM, and focus on the hardware-specific concepts.
% Most instructions may appear in all units, but some restrictions exist:
% Control flow instructions are only legal in functions and processes, time flow instructions only in processes, and instantiations only in entities.

\subsubsection{Hierarchy}
\label{sec:lang_isa_inst}

Hierarchy and structure is described via the \texttt{inst} instruction, which is limited to entities.
The instruction names a process or entity to be instantiated and associates each of its inputs and outputs with a signal (see \cref{fig:lang}ek).

\subsubsection{Signals}
\label{sec:lang_isa_sig}

Signals are created with the \texttt{sig} instruction by providing the type of the value the signal carries, together with its initial value.
The current value of a signal can be probed with the \texttt{prb} instruction, which takes a signal as its argument.
A new value may be driven onto the signal with the \texttt{drv} instruction, which takes a target signal, value to be driven, drive delay, and an optional condition as arguments.
These instructions are limited to processes and entities.
For example in \cref{fig:lang}, the \texttt{@acc\_tb\_initial} process (b) uses \texttt{prb} to ``read'' the value of its input \texttt{\%q} (m), and \texttt{drv} to ``write'' a change to its outputs \texttt{\%en}, \texttt{\%clk}, and \texttt{\%x} (n).
The \texttt{@acc\_tb} entity uses \texttt{sig} to define local signals (h) that connect \texttt{@acc} (k) and \texttt{@acc\_tb\_initial} (e).

\subsubsection{Registers}
\label{sec:lang_isa_reg}

State-holding storage elements such as registers and latches are created with the \texttt{reg} instruction, which is limited to entities.
The \texttt{reg} instruction takes the stored value type and the initial value as first arguments.
These are followed by a list of values, each with a trigger that describes when this value is stored.
The trigger consists of the keyword \texttt{low}, \texttt{high}, \texttt{rise}, \texttt{fall}, and \texttt{both}, followed by a value.
This allows for active-low/high, as well as rising, falling, and dual edge-triggered devices to be described.
To model conditionally-enabled circuits, an optional \texttt{if} gating clause can be used to discard the trigger if some condition is not met.
For example, the optimized \texttt{@acc\_ff} entity in \cref{fig:opt_example}k further ahead uses \texttt{reg} to allocate a rising-edge triggered flip-flop to store the current accumulator state.

\subsubsection{Data Flow}
\label{sec:lang_isa_df}

Data flow instructions, including constants, logic and arithmetic operations, shifts, and comparisons, are a significant part of the instructions in \llhd{}.
For example in \cref{fig:lang}, the \texttt{@acc\_tb\_check} function (a) uses the \texttt{add}, \texttt{mul}, \texttt{div}, and \texttt{eq} instructions to check the accumulator result.
Selection between multiple values is performed with the \texttt{mux} instruction, which takes a sequence of values of the same type as arguments followed by a discriminator that chooses among them.

\subsubsection{Bit-precise Insertion/Extraction}
\label{sec:lang_isa_insext}

Bit-precise control of values is essential to describe digital designs.
The \texttt{insf} and \texttt{extf} instructions allow to set (insert) or get (extract) the value of individual array elements or struct fields.
The \texttt{inss} and \texttt{exts} instructions allow to set or get the value of a slice of array elements or bits of an integer.

\subsubsection{Pointer/Signal Arithmetic}
\label{sec:lang_isa_gep}

The extraction (\texttt{extf} and \texttt{exts}) and shift (\texttt{shl} and \texttt{shr}) instructions can also operate on pointers and signals.
In this mode, these instructions return a new pointer or signal, which directly points at the extracted field or slice or to the shifted value.
These operations are very useful when translating from \glspl{hdl}, where slices or subsets of signals are frequently accessed or driven.
In order to translate into hardware, these partially-accessed signals must be subdivided to the granularity of these accesses, in order to arrive at canonical drive and storage conditions for generated flip-flops or signal wires.

\subsubsection{Control and Time Flow}
\label{sec:lang_isa_cftf}

\emph{Control flow} instructions are the typical ones used in imperative compilers: conditional and unconditional branches, function calls, and returns.
The example in \cref{fig:lang} uses \texttt{br} to implement a loop, \texttt{call} to execute the \texttt{@acc\_tb\_check} function (d), and \texttt{ret} to return from that function.

\emph{Time flow} instructions allow processes to control the passing of time.
The \texttt{wait} instruction suspends execution until one of its operand signals changes and optionally until a certain amount of time has passed.
Execution then resumes at the basic block passed to the \texttt{wait} instruction as its first argument.
The \texttt{halt} instruction suspends execution of the process forever.
The example in \cref{fig:lang} uses \texttt{wait} to suspend execution for \SI{2}{\nano\second} in each loop iteration (f), and \texttt{halt} to stop once the loop terminates (g).

\subsubsection{Memory}
\label{sec:lang_isa_mem}

Stack and heap (or ``dynamic'') memory are required to fully map \glspl{hdl}.
Stack memory holds local variables in functions (e.g., loop variables).
Heap memory is required for Turing completeness, which is necessary to represent \emph{all} simulation and verification code in today's \glspl{hdl}.
For example, SystemVerilog provides builtin dynamic queues and vectors, sparse associative arrays, and a mechanism to call into native code loaded from an object file, or vice versa.
These features require heap allocation and deallocation of memory, and are heavily used in more advanced testbenches and verification code.
Values in allocated memory are loaded and stored with the \texttt{ld} and \texttt{st} instructions.
Stack allocation is implemented by the \texttt{var} instruction, and heap allocation and deallocation by \texttt{alloc} and \texttt{free}, respectively.

\llhd{} code intended for hardware synthesis is expected to require a bounded amount of stack and heap memory known at compile-time, usually none at all.
Bounded heap allocations are guaranteed to be promotable to stack allocations, and bounded stack allocations guarantee that the total amount of required memory is known at compile time.
An algorithm similar to LLVM's memory-to-register promotion allows \llhd{} to promote memory instructions to values and \texttt{phi} nodes.
Lowering to \llhd[2]{} requires \emph{all} stack and heap memory instructions to be promoted in this way, as a design is otherwise not implementable in hardware and can be rejected.
Verification code, which is expected to run in an interpreter, does not have to meet these requirements.

\subsubsection{Intrinsics and Debugging}
\label{sec:lang_isa_dbg}

Additional functionality beyond what is provided as explicit instructions may be represented as intrinsics.
An intrinsic is a call to a predefined function prefixed with ``\texttt{\llhdlower{}.}'' (e.g. \cref{fig:lang}p).
This allows for concepts such as stdin/stdout, file I/O, or assertions to be preserved when transporting from \glspl{hdl} into \llhd{}.
We envision debug information such as \gls{hdl} source locations and naming to be attached to instructions and units via metadata nodes akin to LLVM.
Furthermore, a special \texttt{obs} instruction could be used to describe an observation point for a signal as the user has written it in the original \gls{hdl}.
This allows a design to remain debuggable and recognizable to a user despite aggressive synthesis transformation --- a feature which is currently lacking in commercial synthesizers and place-and-route software.

\section{Mapping HDLs to \llhd{}}
\label{sec:map}

\begin{figure}
    \centering
    \includegraphics[width=\linewidth]{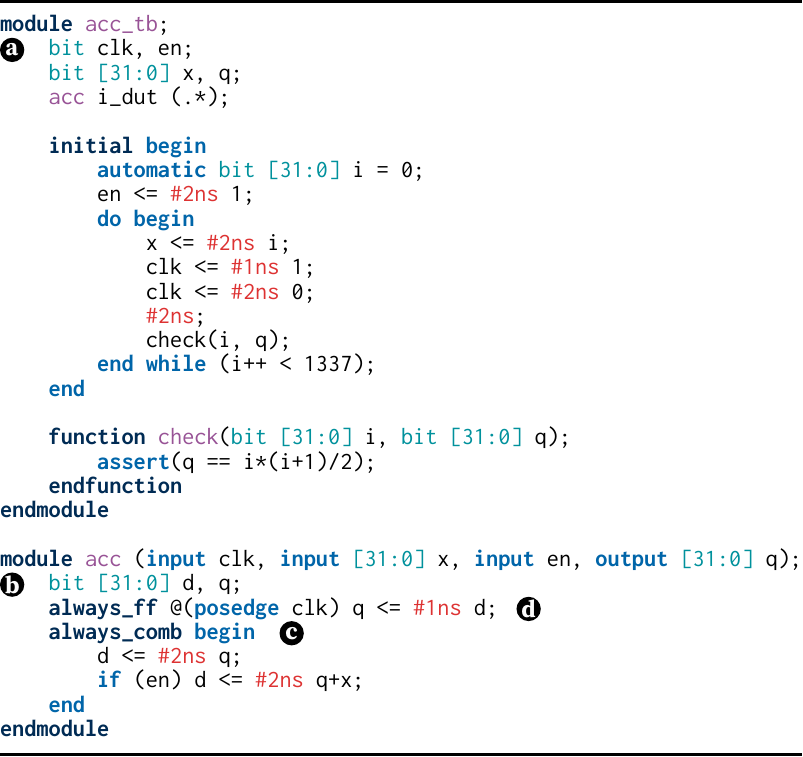}
    \caption{
        SystemVerilog source code for the testbench and accumulator \llhd{} code in \cref{fig:lang} and \cref{fig:opt_example}, as an instructive example as to how \acrshort{hdl} concepts map to \llhd{}.
        See \cref{sec:map} for a detailed description.
    }
    \label{fig:map}
\end{figure}

\llhd{}'s primary goal is to capture designs described in \glspl{hdl} such as SystemVerilog or VHDL.
This explicitly includes simulation and verification constructs and not just the synthesizable subset of a language.
As part of the \llhd{} project we are developing the \emph{\moore{}} compiler, which maps SystemVerilog and VHDL to \llhd{}.
This is comparable to the interaction between Clang and LLVM.
\moore{}'s goal is to map as much of SystemVerilog and VHDL to \llhd{} as possible, providing a reference implementation for both \glspl{hdl} and a platform for future efforts in hardware synthesis, verification, and simulation without the need to reinvent the \gls{hdl} compiler wheel.
This section explores how common constructs in these languages map to \llhd{}, based on our accumulator and testbench running example, the SystemVerilog source code of which is shown in \cref{fig:map}.
The resulting \llhd{} code is shown in \cref{fig:lang} (testbench) and \cref{fig:opt_example} (accumulator).

%-------------------------------------------------------------------------------
\subsection{Hierarchy}

The fundamental elements of reuse and hierarchy are ``modules'' in SystemVerilog and ``entities'' in VHDL.
Both describe a circuit as a list of input and output ports and a body of signals, sub-circuits, and processes.
Thus, they trivially map to \llhd{} entities, for example \cref{fig:map}a to \cref{fig:lang}c, or \cref{fig:map}b to \cref{fig:opt_example}m.

%-------------------------------------------------------------------------------
\subsection{Processes}
\label{sec:map_procs}

Processes are the main circuit modeling tool in most \glspl{hdl}.
SystemVerilog provides the \texttt{always}, \texttt{always\_ff}, \texttt{always\_latch}, \texttt{always\_comb}, \texttt{initial}, and \texttt{final} constructs, while VHDL has a general-purpose \texttt{process} concept.
Using these constructs, a circuit can be described in a \emph{behavioural} fashion by providing an imperative subprogram that maps a change in input signals to a change in output signals: essentially, the body of a process is re-executed whenever one of its input signals changes.
\llhd{}'s processes are designed to capture these constructs through an almost verbatim translation from an \gls{hdl} process.
Processes allow for a very high-level and general description of circuits.
However, it is common practice to follow a strict modeling style in \glspl{hdl} to ensure synthesizers infer the proper hardware, which we briefly describe in the following.

\subsubsection{Combinational Processes}
\label{sec:map_procs_comb}

Combinational processes describe a purely functional mapping from input signals to output signals, without any state-keeping elements such as flip-flops as side effect.
Synthesizers can readily map such processes to logic gates.
A process is combinational if there are no control flow paths that leave any of its output signals unassigned.
Combinational processes can be mapped to a pure data flow graph of logic gates.
Consider the \texttt{always\_comb} process in \cref{fig:map}c, which directly maps signals \texttt{q} and \texttt{x} to an output value \texttt{d}.
This translates into the \texttt{@acc\_comb} \llhd{} process in \cref{fig:opt_example}n.

\subsubsection{Sequential Processes}
\label{sec:map_procs_seq}

Sequential processes describe state-keeping elements such as flip-flops and latches.
A process is sequential if at least some of its output signals are only assigned under certain conditions.
Synthesizers detect these kinds of processes, usually requiring the designer to adhere to a very strict pattern, and map them to the corresponding storage gate.
Consider the \texttt{always\_ff} process in \cref{fig:map}d, which maps to the \texttt{@acc\_ff} \llhd{} process in \cref{fig:opt_example}p.
\llhd{} can capture register descriptions in the behavioral form (see \cref{fig:opt_example}p) but also provides an explicit \texttt{reg} instruction (see \cref{fig:opt_example}k) to canonically represent registers, which is inferred by lowering passes discussed in \cref{sec:opt}.

\subsubsection{Mixed Processes}

SystemVerilog and VHDL allow designers to mix combinational and sequential styles in one process, but support for this is limited even in commercial synthesizers.
The desequentialization pass (\cref{sec:opt_deseq}) can help split mixed processes into combinational and sequential parts for higher compatibility with many synthesizers.

%-------------------------------------------------------------------------------
\subsection{Generate Statements and Parameters}

Many \glspl{hdl} provide \emph{generate statements} and \emph{parameters} to allow for parametrized generation of hardware.
\llhd{} does \emph{not} provide such constructs, but rather expects these statements to be unrolled already by the compiler frontend (e.g., \moore{}).
The rationale is that \gls{hdl} designs parametrized over constants and types lead to significant changes in the generated hardware that go beyond mere type or constant substitution.
For example, a parameter might cause the generate statements in a design to unroll to a completely different circuit.
Capturing this flexibility in \llhd{} would require a significant meta-programming layer to be added, which in our opinion is best left to a higher-level language or \gls{ir}.

%-------------------------------------------------------------------------------
\subsection{Verification}

Modern \glspl{hdl} feature constructs to verify hardware designs.
SystemVerilog, for example, provides the \texttt{assert}, \texttt{assume}, and \texttt{require} constructs.
We propose to map these to \llhd{} intrinsics such as \texttt{\llhdlower{}.assert}.
A simulator may then choose to emit error messages when the condition passed to such an intrinsic is false.
A formal verification tool, on the other hand, can extract these intrinsics and set up a satisfiability problem or perform bounded model checking~\cite{clarke2001bounded}.
Higher-level constructs to describe \gls{ltl} and \gls{ctl} properties~\cite{gupta1992formal} shall be mapped to intrinsics as well, such that a formal verification tool can recover them from the \gls{ir}.
An interesting side-effect of preserving these verification constructs as intrinsics is that \gls{fpga} mappings of an \llhd{} design may choose to implement the assertions in hardware, to perform run-time checks of a circuit.
These features are not yet implemented in our research prototype, and an \llhd{}-based verification tool is yet to be written.

\section{Lowering to \llhd[2]{}}
\label{sec:opt}

\begin{figure}
    \centering
    \includegraphics[width=\columnwidth]{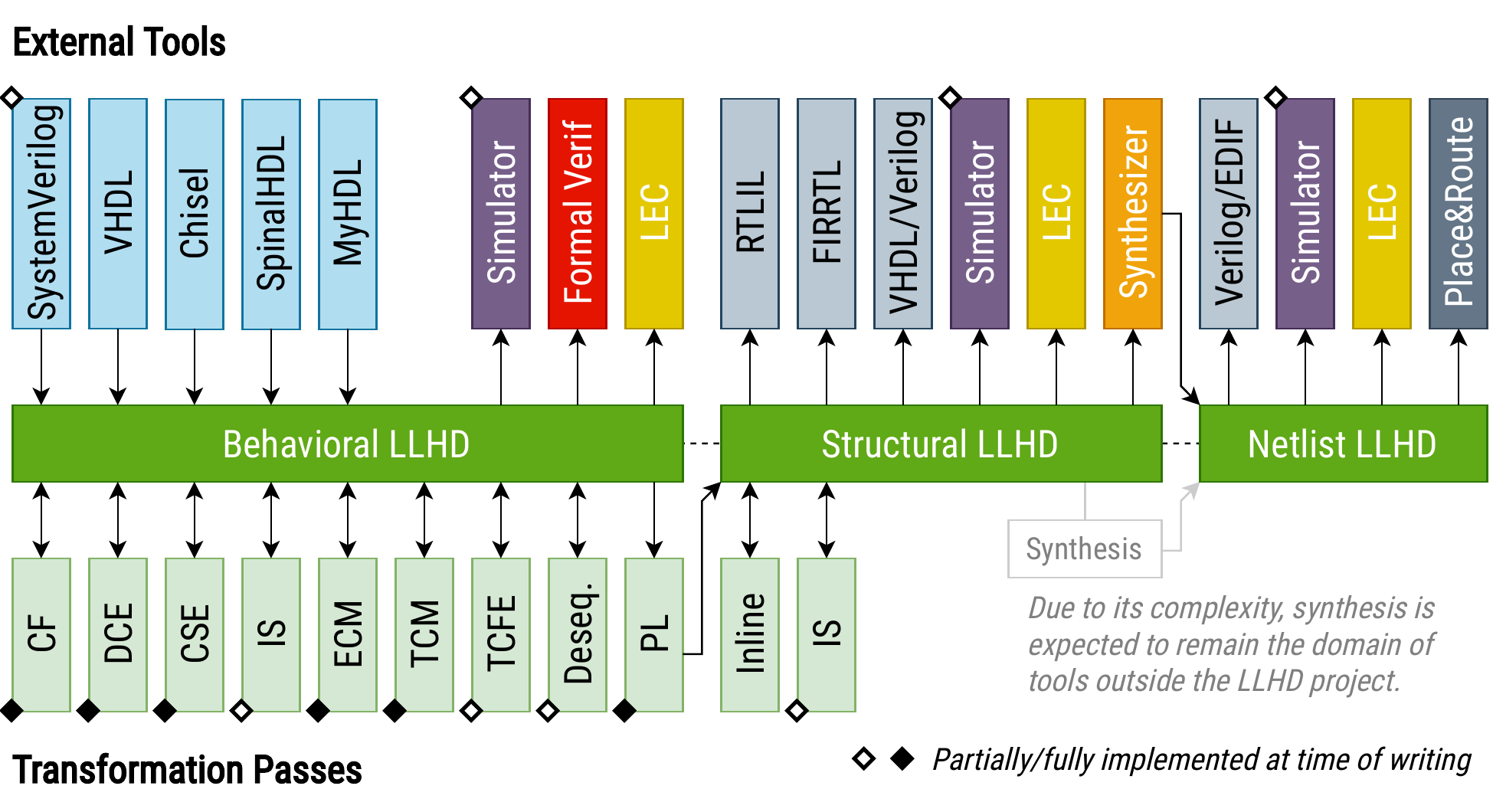}
    \caption{
        Optimization and transformation passes on the different \gls{ir} levels of \llhd{}.
        See \cref{sec:opt} for a detailed description.
    }
    \label{fig:opt_flow}
\end{figure}

\begin{figure*}
    \centering
    \includegraphics[width=\linewidth]{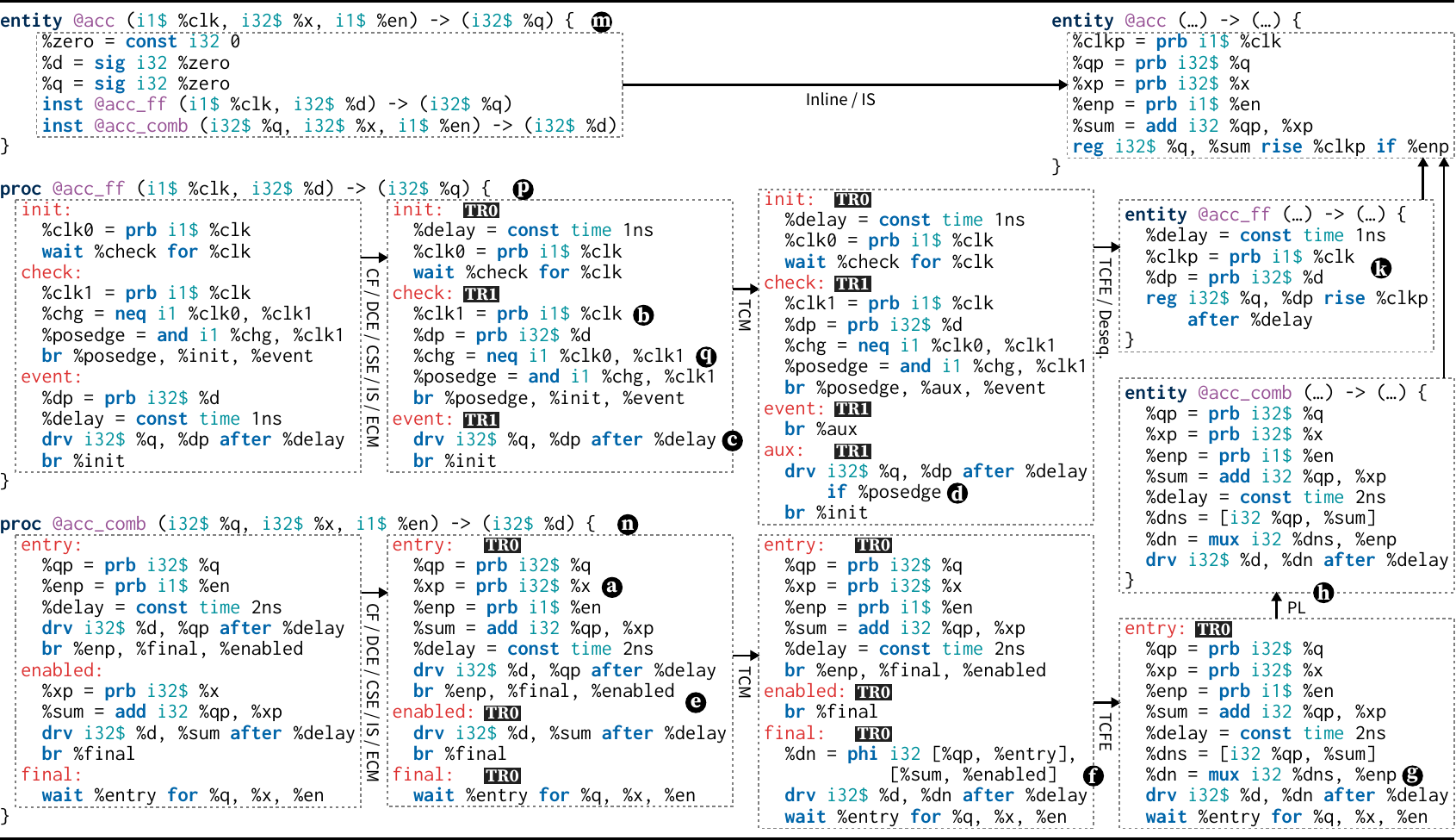}
    \caption{
        An end-to-end example of how a simple accumulator design is lowered from \llhd[1]{} (left) to \llhd[2]{} (right).
        See \cref{sec:opt} for a detailed description of the individual steps.
        The processes \texttt{@acc\_ff} and \texttt{@acc\_comb} are lowered to entities through various transformations, and are eventually inlined into the \texttt{@acc} entity.
    }
    \label{fig:opt_example}
\end{figure*}

One of the key contributions of \llhd{} is a framework to translate the high-level behavioural circuit descriptions of an \gls{hdl} into a lower-level structural description that can be readily synthesized.
Existing commercial and open-source synthesizers all have redundant proprietary implementations of this procedure; see \cref{fig:intro}.
With \llhd{} and language frontends such as \moore{}, this translation can be performed as a lowering pass on the \gls{ir} directly, rather than individually by each synthesizer.
The general objective of this lowering is to replace branches with branch-free code.
In particular, it consists of the following high-level steps:
\begin{itemize}
    \item Reduce complexity of operations (\cref{sec:opt_basic})
    \item Move arithmetic out of basic blocks (\acrshort{ecm}, \cref{sec:opt_ecm})
    \item Move drives out of basic blocks (\acrshort{tcm}, \cref{sec:opt_tcm})
    \item Replace \texttt{phi} and control flow with \texttt{mux} (\acrshort{tcfe}, \cref{sec:opt_tcfe})
    \item Replace trivial processes with entities (\acrshort{pl}, \cref{sec:opt_pl})
    \item Identify flip-flops and latches (Deseq., \cref{sec:opt_deseq})
\end{itemize}
Consider the accumulator design in \cref{fig:opt_example}, which is a typical example of \llhd[1]{} as it is generated from a SystemVerilog source.
Let us step through the transformations in \cref{fig:opt_flow} to lower this example to \llhd[2]{}.

%-------------------------------------------------------------------------------
\subsection{Basic Transformations}
\label{sec:opt_basic}

In a first step, we apply basic transformations such as \gls{cf}, \gls{dce}, and \gls{cse}, which are equivalent to their LLVM counterparts.
Furthermore, \gls{is} is used as a peephole optimization to reduce short instruction sequences to a simpler form, similar to LLVM's instruction combining.
To facilitate later transformations, all function calls are inlined and loops are unrolled at this point.
Where this is not possible, the process is rejected.

%-------------------------------------------------------------------------------
\subsection{\glsreset{ecm}\gls{ecm}}
\label{sec:opt_ecm}

\Gls{ecm} moves instructions ``up'' in the \gls{cfg}, to facilitate later control flow elimination.
This is similar to and subsumes \gls{licm} in LLVM in that code is hoisted into predecessor blocks, but \gls{ecm} does this in an eager fashion.
The underlying motif of lowering to \llhd[2]{} is to eliminate control flow, since that does not have an equivalent in hardware.
An essential step towards this is to \emph{eagerly} move instructions into predecessor blocks as far as possible.
As shown in \cref{fig:opt_example}a, this moves all constants into the entry block, and arithmetic instructions to the earliest point where all operands are available.
Special care is required for \texttt{prb} as in \cref{fig:opt_example}b, which \emph{must not} be moved across \texttt{wait}, as that would imply a semantic change.

%-------------------------------------------------------------------------------
\subsection{\glsreset{tcm}\gls{tcm}}
\label{sec:opt_tcm}

The \texttt{wait} instructions naturally subdivide a process into different \glspl{tr}, i.e. sections of code that execute during a fixed point in physical time.
A key step towards eliminating control flow is to move \texttt{drv} instructions into a single exiting block for their respective \gls{tr}.
The condition under which the control flow reaches a \texttt{drv} instruction before the move is added as an additional operand to the instruction.
% Intuitively speaking, this translates the condition from the \gls{cfg} into the \gls{dfg}.
Let us elaborate in more detail by first introducing the concept of \glspl{tr}.

\subsubsection{\glsreset{tr}\gls{tr}}
\label{sec:opt_tcm_tr}

The capability of \texttt{wait} to suspend execution mid-process until a later point in time calls for novel techniques to reason about the synchronicity of instructions.
More specifically, we would like to know if two instructions execute at the same instant of physical time \emph{under all circumstances}.
Consider the \texttt{\%clk} signal in \cref{fig:opt_example}q as an illustrative example:
when we reach the \texttt{neq} instruction, we would like to be able to reason that \texttt{\%clk0} is an ``old'' sampling of \texttt{\%clk} from before the \texttt{wait}, and that \texttt{\%clk1} reflects the current state of \texttt{\%clk}.
Each basic block in \llhd{} has an associated \gls{tr}.
Multiple blocks may belong to the same \gls{tr}.
The set of blocks in the same \gls{tr} represents the bounds within which \texttt{prb} and \texttt{drv} instructions may be rearranged without changing the process behaviour.
As an intuition, \glspl{tr} are assigned to individual blocks based on the following rules:
\begin{enumerate}
    \item If any predecessor has a \texttt{wait} terminator, or this is the entry block, generate a \emph{new} \gls{tr}.
    \item If all predecessors have the \emph{same} \gls{tr}, inherit that \gls{tr}.
    \item If they have \emph{distinct} \glspl{tr}, generate a \emph{new} \gls{tr}.
\end{enumerate}
Note that as a result of rule 3, there is one unique entry block for each \gls{tr} where control transfers to from other \glspl{tr}.
\cref{fig:opt_example}ab shows the temporal regions assigned to the individual blocks.
Note that the flip-flop process \texttt{@acc\_ff} has two \glspl{tr}, whereas the combinational process \texttt{@acc\_comb} has just one.

\subsubsection{Single Exiting Block per \gls{tr}}
\label{sec:opt_tcm_single_exit}

We would like each \gls{tr} to have a single exiting block.
This is essential to have a single point in the \gls{cfg} to move \texttt{drv}s to such that they are always executed in this \gls{tr}.
If \gls{tr} A has multiple control flow arcs leading to \gls{tr} B, an additional intermediate block is inserted in order to have a single arc from \gls{tr} A to B.
This is always possible since as a result of rule 3 in \cref{sec:opt_tcm_tr}, all branches to \gls{tr} B target the same unique entry block.
In \cref{fig:opt_example}b for example, blocks \texttt{check} and \texttt{event} both branch to \texttt{init}, which is in a different \gls{tr}.
In this case an auxiliary block is created as part of \gls{tcm}, where \texttt{check} and \texttt{event} branch to.

\subsubsection{Moving Drive Instructions}
\label{sec:opt_tcm_move_drv}

As the main part of \gls{tcm}, \texttt{drv} instructions are moved into the single exiting block of their \gls{tr}.
We first find the closest common dominator of the exiting block and the instruction.
If no such dominator exists, the instruction is left untouched, which later causes the lowering to reject the process.
As a second step, we find the sequence of branch decisions that cause control to flow from the dominator to the \texttt{drv} instruction.
This essentially builds a chain of \texttt{and}s with the branch conditions ``along the way'', or their inverse, as operands.
As the final third step, the \texttt{drv} instruction is moved into the exiting block, and the expression found in the second step is set as the \texttt{drv}'s optional condition operand.
In \cref{fig:opt_example}c, control flow only reaches the \texttt{drv} if the \texttt{\%posedge} branch argument is true.
Consequently, \texttt{\%posedge} is added as drive condition in \cref{fig:opt_example}d.
In \cref{fig:opt_example}e, control flow always reaches the \texttt{drv}s, which are consequently moved into the existing single exiting block \texttt{final} without adding a condition operand, as in \cref{fig:opt_example}f.
Since both \texttt{drv}s target the same signal, they are coalesced into one instruction, and selection of the driven value is factored out into a \texttt{phi} instruction.

%-------------------------------------------------------------------------------
\subsection{\glsreset{tcfe}\gls{tcfe}}
\label{sec:opt_tcfe}

The goal now is to replace control flow with data flow, branches with multiplexers.
The previous transformations leave many empty blocks behind.
\gls{tcfe} eliminates these blocks such that only one block remains per \gls{tr}.
This is the case in \cref{fig:opt_example}df, where only the \texttt{init}, \texttt{check}, and \texttt{entry} blocks remain.
Furthermore, all \texttt{phi} instructions are replaced with \texttt{mux} instructions, as shown in \cref{fig:opt_example}g.
The selector for the \texttt{mux} instruction is found in the same way as the \texttt{drv} condition in \cref{sec:opt_tcm_move_drv}.
As a result, combinational processes (\cref{sec:map_procs_comb}) now consist of a single block and \gls{tr}, and sequential processes (\cref{sec:map_procs_seq}) of two blocks and \glspl{tr}.
Processes for which neither holds are rejected by the lowering.

%-------------------------------------------------------------------------------
\subsection{\glsreset{pl}\gls{pl}}
\label{sec:opt_pl}

At this point, processes with a single block and a \texttt{wait} terminator of the correct form are lowered to an entity.
This is done by removing the \texttt{wait} and moving all other instructions to an entity with the same signature.
In order for this to be equivalent, the \texttt{wait} must be sensitive to all \texttt{prb}'d signals.
See \cref{fig:opt_example}h for an example where this is the case.

%-------------------------------------------------------------------------------
\subsection{Desequentialization (Deseq.)}
\label{sec:opt_deseq}

For the remaining processes we would like to identify if they describe a sequential circuit such as a flip-flop or latch.
\Glspl{hdl} expect these to be inferred from signals that are only driven under certain conditions, e.g., if a clock signal changed or a gate signal is high.
The \gls{tcm} pass canonicalizes processes into a form which makes this rather straightforward.
We only consider processes with two basic blocks and \glspl{tr}, which covers all relevant practical \gls{hdl} inputs.
In a first step, we canonicalize the condition operand of each \texttt{drv} into its \gls{dnf}.
The \gls{dnf} exists for all boolean expressions, is trivially extended to \texttt{eq} and \texttt{neq}, and can retain all non-canonicalizable instructions as opaque terms.
Each separate disjunctive term of the \gls{dnf} identifies a separate trigger for the flip-flop or latch.
The \texttt{drv} in \cref{fig:opt_example}d, for example, has the canonical condition $\neg\text{\texttt{\%clk0}} \land \text{\texttt{\%clk1}}$.
In a second step, we identify which terms of the condition are sampled \emph{before} the \texttt{wait}, and which are sampled \emph{after}.
This is done based on the \gls{tr} of the corresponding \texttt{prb}.
The \gls{tr} of the \texttt{wait} is considered the ``past'', and the \gls{tr} of the \texttt{drv} is considered the ``present''.
In a third step, we isolate terms $T$ which are sampled both in the past ($T_0$) and the present ($T_1$), and pattern match as follows:
\begin{itemize}
    \item $\neg T_0 \land T_1$ is a \emph{rising} edge on $T$
    \item $T_0 \land \neg T_1$ is a \emph{falling} edge on $T$
    \item $(\neg T_0 \land T_1) \lor (T_0 \land \neg T_1)$ are \emph{either} edges on $T$
\end{itemize}
All other terms are moved into the set of ``trigger conditions''.
At this point, a separate entity is created which will hold the identified sequential elements.
In a final step, all \texttt{drv}s, for which the above trigger identification was successful, are mapped to an equivalent \texttt{reg} instruction.
Each trigger is added to the \texttt{reg} separately, together with the corresponding set of trigger conditions: edge terms are mapped to corresponding \texttt{rise}, \texttt{fall}, or \texttt{both} edge triggers, and all remaining terms to \texttt{high} or \texttt{low} level triggers.
Furthermore, the entire \gls{dfg} of the \texttt{drv} operands, namely driven signal, value, delay, and condition, is added to the entity.
See \cref{fig:opt_example}k.
This procedure identifies and isolates edge- and level-triggered sequential elements into an entity.
The remaining process is either empty and removed, lowered by \gls{pl}, or rejected.

%-------------------------------------------------------------------------------
\subsection{Synthesizability}
\label{sec:opt_synth}

The class of ``synthesizable'' hardware descriptions and subsets of \glspl{hdl} is defined rather loosely.
In practice, hardware descriptions adhere to sufficiently strict coding guidelines that enable the above transformations to occur.

\section{Toolflow Integration}
\label{sec:flow}

Languages such as SystemVerilog are too high-level to reliably transport digital circuits through the design flow.
This is mainly due to the difficulty to consistently implement these languages, as described in \cref{sec:intro}.
For example, designs are simulated and verified in their \gls{hdl} description, by a simulator or verification tool.
A \gls{lec} is then used to verify that the design has been mapped correctly from \gls{hdl} to an equivalent netlist.
Not only does this require the \gls{lec} to fully implement the SystemVerilog standard as well, it also only verifies that the synthesizer's interpretation of the \gls{hdl} matches the \gls{lec}'s.
However, it does not verify whether the simulation and verification tool's interpretation matches the \gls{lec}'s.
This is potentially disastrous given the complexity of languages such as SystemVerilog.

\llhd{} provides a design representation that is much simpler to implement consistently.
Simulating and verifying the \llhd{} mapping of a circuit rather than its original \gls{hdl} means converging to a single simple representation early in the design flow.
Synthesis tools and \glspl{lec} running on this reference \llhd{} mapping then ensure correct translation of the circuit into a netlist.
\llhd{}'s simplicity offers a much smaller ``surface for implementation errors''.

Many commercial tools already use a proprietary \gls{ir} internally.
These are generally accessible to the user, such that a \llhd[2]{} description can be mapped directly to such a tool's \gls{ir}.
Where this is not possible, the description may be mapped to a simple, structural Verilog equivalent to be ingested by the tool.
Ideally, vendors eventually support direct input of \llhd{}, but this is not required.

We conclude that \llhd{} fits very well into existing commercial tool flows.
Its simplicity makes it a prime candidate to harden the \gls{hdl}-to-netlist verification chain.
Furthermore, its expressivity allows it to subsume other \glspl{ir}, offering a platform to transport designs between FIRRTL~\cite{izraelevitz2017reusability}, RTLIL~\cite{wolf2018rtlil}, CoreIR~\cite{mattarei2018cosa}, and others.

Moreover, \llhd{} significantly lowers the hurdle for innovation in the digital hardware design ecosystem.
For instance, a new \gls{hdl} only needs to be lowered to \llhd{} to become supported by all existing toolchains, \gls{hls} compilers can generate \llhd{} as output, simulators only have to parse the simple \llhd{} rather than complex \glspl{hdl}, and synthesizers can take \llhd[2] as their input.

\section{Evaluation}
\label{sec:eval}

We have implemented the \moore{} compiler and \llhd{} as a vertical research prototype.
\moore{} covers enough of the SystemVerilog standard to show merit on a non-trivial set of open-source hardware designs:
specifically, we consider the designs listed in \cref{tab:eval_sim}, which range from simple arithmetic primitives, over \gls{fifo} queues, \glspl{cdc}, and data flow blocks, up to a full RISC-V processor core~\cite{waterman2014risc}.
The lines of SystemVerilog code (``LoC'') in \cref{tab:eval_sim} provides a rough indication of the respective hardware complexity.
The designs are mapped from SystemVerilog to \llhd[1]{} with the \moore{} compiler, without any optimizations enabled.

\begin{table}
\caption{
    Evaluation of the simulation performance of \llhd{}.
    We compare the \llhd{} reference interpreter to a JIT-accelerated \llhd{} simulator and to a commercial simulator.
    The former two operate on unoptimized \llhd{} code as emitted by the \moore{} frontend with the \texttt{-O0} flag.
    We list the lines of code ``LoC'' and executed clock ``cycles'' to provide an indication for the design and simulation complexity.
    Traces match between the two simulators for all designs.
    See \secref{sec:eval_sim} for a detailed description.
}
\begin{threeparttable}
\begin{tabularx}{\linewidth}{@{}Lrrrrr@{}}
    \toprule
    &&&
    \multicolumn{3}{c@{}}{\textbf{Sim.\ Time} [s]} \\
    \cmidrule(l){4-6}
    \textbf{Design} &
    \textbf{LoC} &
    \textbf{Cycles} &
    Int.\tnote{1} &
    JIT\tnote{2} &
    Comm.\tnote{3} \\
    \midrule
    Gray Enc./Dec.   &   17 & 12.6\,M &  9740 &  6.33 &  6.07 \\
    FIR Filter       &   20 &    5\,M &  4430 & 10.35 & 14.60 \\
    LFSR             &   30 &   10\,M &  2350 & 14.53 & 14.10 \\
    Leading Zero C.  &   52 &    1\,M & 11000 &  3.23 &  7.84 \\
    FIFO Queue       &  102 &    1\,M &  1370 &  5.92 &  5.55 \\
    CDC (Gray)       &  108 &    1\,M &  1380 &  8.72 &  6.45 \\
    CDC (strobe)     &  122 &  3.5\,M &  1570 &  9.39 &  6.11 \\
    RR Arbiter       &  159 &    5\,M & 49400 & 10.92 & 25.54 \\
    Stream Delayer   &  219 &  2.5\,M &   477 &  4.28 &  4.99 \\
    RISC-V Core      & 3479 &    1\,M & 24000 & 23.44 &  4.47 \\
    \bottomrule
\end{tabularx}
\begin{tablenotes}
    \item[1] \llhd{} reference interpreter (\emph{\llhdsim{}}), extrapolated;
    \item[2] JIT-accelerated simulator (\emph{\llhdblaze{}});
    \item[3] Commercial HDL simulator
\end{tablenotes}
\end{threeparttable}
\label{tab:eval_sim}
\end{table}

\begin{table}
\caption{
    Comparison against other hardware-targeted intermediate representations.
    Most other \glspl{ir} are geared towards synthesis and the resulting netlists.
    See \secref{sec:eval_irs} for a detailed description.
}
\setlength{\tabcolsep}{.45em}
\begin{threeparttable}
\begin{tabularx}{\linewidth}{@{}Llccccccc@{}}
    \toprule
    % {} & {} & {} & {} & \multicolumn{3}{c}{Can represent} \\
    \textbf{IR} &
    \rot{\textbf{No.\ of Levels}} &
    \rot{\parbox{5em}{\textbf{Turing-Complete}}} &
    \rot{\parbox{7em}{\textbf{Verification}\\[-.3em]{\small(e.g., assertions)}}} &
    \rot{\parbox{7em}{\textbf{9-Valued Logic}\\[-.3em]{\small(IEEE 1164~\cite{ieee1164})}}} &
    \rot{\parbox{7em}{\textbf{4-Valued Logic}\\[-.3em]{\small(IEEE 1364~\cite{ieee1364})}}} &
    \rot{\textbf{Behavioral}} &
    \rot{\textbf{Structural}} &
    \rot{\textbf{Netlist}} \\
    \midrule
    \llhd{} [us]                             & 3               & \checkmark   & \checkmark & \checkmark & \checkmark & \checkmark & \checkmark & \checkmark \\
    FIRRTL \cite{izraelevitz2017reusability} & 3\tnote{$\dag$} & --           & --         & --         & --         & --         & \checkmark & \checkmark \\
    CoreIR \cite{mattarei2018cosa}           & 1               & --           & \checkmark & --         & --         & --         & \checkmark & --         \\
    µIR \cite{sharifian2019muir}             & 1               & --           & --         & --         & --         & --         & \checkmark & --         \\
    RTLIL \cite{wolf2018rtlil}               & 1               & --           & --         & --         & \checkmark & \checkmark & \checkmark & --         \\
    LNAST \cite{wang2019lnast}               & 1               & --           & --         & --         & --         & \checkmark & --         & --         \\
    LGraph \cite{wang2019lgraph}             & 1               & --           & --         & --         & --         & --         & \checkmark & \checkmark \\
    netlistDB \cite{anon2019netlistdb}       & 1               & --           & --         & --         & --         & --         & \checkmark & \checkmark \\
    \bottomrule
\end{tabularx}
\begin{tablenotes}
    \item[$\dag$] Mentioned conceptually but not defined precisely
\end{tablenotes}
\end{threeparttable}
\label{tab:eval_irs}
\end{table}

%-------------------------------------------------------------------------------
\subsection{Circuit Simulation}
\label{sec:eval_sim}

In a first evaluation, we simulate the designs in \cref{tab:eval_sim} mapped to \llhd{} using \emph{\llhdsim{}}, the \llhd{} reference interpreter, and \emph{\llhdblaze{}}, a simulator leveraging \gls{jit} code generation.
Both simulators support all levels of the \llhd{} \gls{ir}.
\llhdsim{} is deliberately designed to be the simplest possible simulator of the \llhd{} instruction set, rather than the fastest.
\llhdblaze{} is a first implementation of \gls{jit} compilation to show the potential of massively accelerated simulation.
For each of the designs, we compare the execution time of a commercial SystemVerilog simulator (``Comm.'') with our two approaches in \cref{tab:eval_sim}.
Simulations were executed on an Intel Core i7-4770 CPU running at \SI{3.4}{\GHz}.
Without \gls{jit} compilation, \llhdsim{} is slower than its commercial counterpart, which trades initial optimization overhead at startup for increased simulation performance.
Most importantly, however, the \llhd{} simulation trace is equal to the one generated by the commercial simulator for all designs.
That is, even with a simple prototype implementation of \moore{} and \llhd{}, we can fully and correctly represent a RISC-V processor in \llhd[1]{}.

\Gls{jit} compilation provides an opportunity to massively accelerate \llhd{} simulation.
For this, we map \llhd{} to LLVM \gls{ir} and then use LLVM to optimize for execution on the simulation machine.
This makes \llhdblaze{} (``JIT'') competitive with commercial simulators, even though the former is a prototype whereas the latter have been optimized for decades.
In some cases, \llhdblaze{} is already up to $2.4 \times$ faster than commercial simulators.

These benefits manifest themselves even on entirely \emph{unoptimized} \llhd{} code as it is emitted by \moore{} with the \texttt{-O0} flag, comparable to Clang's equivalent.
We expect the discrepancies between \llhdblaze{} and the commercial simulator to disappear as we add optimizations to the simulator in the future.
This will especially affect complex benchmarks such as the RISC-V core, where the lack of code optimization currently incurs significant overheads.
Note that the lines of SystemVerilog code and simulation cycles do not always fully portray the complexity of a design:
some SystemVerilog constructs can produce significantly more \llhd{} code than others, leading to significant differences in simulation time.

Much of the complexity and engineering effort of our work lies in \moore{}, the \gls{hdl} compiler frontend, which tackles the extensive SystemVerilog standard.
Further extensions of \moore{} to implement additional SystemVerilog constructs can vastly increase the scope of designs that can be mapped to \llhd{}.
Once a design is in \llhd{}, simulating it correctly is trivial.

%-------------------------------------------------------------------------------
\subsection{Comparison with other Hardware IRs}
\label{sec:eval_irs}

\llhd{} is not the first \gls{ir} targeted at hardware design.
However, to our knowledge it is the first multi-level \gls{ir} capable of capturing a design throughout the entire hardware design flow, from simulation to synthesized netlist.
\cref{tab:eval_irs} compares the part of the design flow covered by \llhd{} and other \glspl{ir}.
We observe that almost all other \glspl{ir} support structural descriptions of circuits and many \glspl{ir} support representing a synthesized netlist.
This is due to the fact that most of them were designed as synthesis \glspl{ir} to capture and apply transformations before and after synthesis.
A notable example is FIRRTL, which acts as an interface between the Chisel~\cite{bachrach2012chisel} frontend and subsequent synthesis.
As the only other \gls{ir}, FIRRTL defines multiple levels of abstraction together with transformations to lower designs from higher to lower levels.
µIR is designed to capture entire accelerator architectures as structural description and caters more to an \gls{hls} flow.
RTLIL and LNAST both support the behavioural description of circuits.
RTLIL is geared towards simplifying Verilog input to make it amenable for synthesis.
LNAST represents behavioural \gls{hdl} code as language-agnostic \gls{ast} and offers transformation passes.
As the only other \gls{ir}, CoreIR focuses on adding formal verification support to \glspl{hdl}.
No other \gls{ir} is Turing-complete, which is essential to represent arbitrary reference models and verification constructs.
FIRRTL provides limited support for testbench constructs in the form of message logging and clocking control.

\llhd{} covers all stages of the flow:
\emph{\llhd[1]{}} captures testbench, verification, and behavioural circuit descriptions, to facilitate interfacing with \glspl{hdl};
\emph{\llhd[2]{}} captures circuits structurally, to interface with synthesis and low-level circuit analysis and transformation tools;
and \emph{\llhd[3]{}} represents the individual gates of a finalized circuit.
Hence \llhd{} is the only \gls{ir} capable of representing the full semantics of the SystemVerilog/VHDL input language for the full hardware design flow.

%-------------------------------------------------------------------------------
\subsection{Size Efficiency}
\label{sec:eval_size}

\begin{table}
\caption{
    Size efficiency of the human-readable text representation, an estimate of a prospective bitcode, and the in-memory data structures of \llhd{}.
    See \cref{sec:eval_size} for details.
}
\begin{threeparttable}
\begin{tabularx}{\linewidth}{@{} L rrrr @{}}
    \toprule
    & \textbf{SV} & \multicolumn{3}{c@{}}{\textbf{\llhd{}} [\si{\kilo\byte}]} \\
    \cmidrule(l){3-5}
    \textbf{Design} & [\si{\kilo\byte}] & Text & Bitcode\tnote{1} & In-Mem. \\
    \midrule
    Gray Enc./Dec.  &   3.0 &  11.9 &  3.6 &   41.8 \\
    FIR Filter      &   1.2 &  12.9 &  3.8 &   46.7 \\
    LFSR            &   2.4 &   4.9 &  1.8 &   18.4 \\
    Leading Zero C. &   4.6 &  97.4 & 21.8 &  309.1 \\
    FIFO Queue      &   7.1 &  20.5 &  6.5 &   71.3 \\
    CDC (Gray)      &   8.5 &  38.1 & 12.5 &  129.5 \\
    CDC (strobe)    &   6.3 &  17.7 &  6.0 &   63.3 \\
    RR Arbiter      &  11.2 &  39.4 &  9.9 &  129.8 \\
    Stream Delayer  &  12.2 &  18.7 &  6.7 &   68.6 \\
    RISC-V Core     & 174.4 & 349.4 & 93.6 & 1096.3 \\
    \bottomrule
\end{tabularx}
\begin{tablenotes}
    \item[1] estimated
\end{tablenotes}
\end{threeparttable}
\label{tab:eval_size}
\end{table}

An important aspect of a hardware \gls{ir} is how efficiently it captures a design in terms of memory used, both in on-disk and in-memory representations.
\cref{tab:eval_size} shows the size in \si{\kilo\byte} occupied by the previously introduced designs.

For on-disk representations, we consider the SystemVerilog \gls{hdl} code as a baseline (first numeric column).
First, we observe that the \emph{unoptimized} \llhd{} text (second column) emitted by the \moore{} compiler using the \texttt{-O0} flag (equivalent to the corresponding Clang/GCC flag) is significantly larger than SystemVerilog.
This is due to the fact that \moore{} tries to keep code generation as simple as possible and emits numerous redundant operations.
Furthermore, many operations which are implicit in SystemVerilog, for example, expression type casts and value truncation/extension, require explicit annotation in \llhd{}.
Future optimizations can reduce redundant operations and produce more compact instructions, which we expect to be significantly smaller.
Utilizing a binary ``bitcode'' representation (third column) instead can greatly reduce the on-disk size of the design, such that compiled \gls{hdl} input represented as \llhd{} unoptimized bitcode is already comparable in size to the input source code.
The bitcode itself is not yet implemented.
Sizes are estimated based on a strategy similar to LLVM's bitcode, considering techniques such as run-length encoding for numbers, interning of strings and types, compact encodings for frequently-used primitive types and value references.
This makes \llhd{} a viable format to transport a design into the initial stages of the design flow, such as testbenches and simulation, formal verification, and synthesis.

The in-memory size of an \gls{ir} (last column of \cref{tab:eval_size}) is even more important for transformation and optimization passes.
While there is no baseline for this, we observe that even a full RISC-V core only requires \SI{1}{\mega\byte} of memory.
As the in-memory complexity scales linearly with the complexity of the design, we argue that representing even entire \gls{soc} designs with hundreds of CPU cores will be feasible with tens of gigabytes of memory, which is fully viable for today's workstations.

\section{Related Work}
\label{sec:relwork}

Intermediate representations are an established and very successful concept both for representing imperative programs and the design of hardware circuits.
This work has been in development over the past three years and predates efforts such as MLIR \cite{lattner2020mlir}, which aims at providing a unifying framework for defining compiler \glspl{ir}.
The proposed concepts can likely be expressed in MLIR.

%-------------------------------------------------------------------------------
\subsection{Compiler Intermediate Representations}
\label{sec:relwork_llvm}

Since the early days of compilers and program optimizations it has been clear that in the process of translating a computer program to machine code the program passes a variety of different representations~\cite{rustin1972design}.
Machine-independent internal program representations have been discussed as early as 1979 in the PQCC project~\cite{Cattell:1979:CGM:800229.806955, schatz1979tcol} which introduced a tree-structured common optimization language (TCOL) and showed its use in an optimizing Ada compiler.
Intermediate representations are standard in most compilers today.
Functional programming languages often use continuation-passing style representations~\cite{steele1976lambda}.
For imperative programming languages, SSA-based intermediate representations~\cite{alpern1988detecting, cytron1989efficient} have shown most successful, and are used in large compiler infrastructures such as LLVM~\cite{lattner2004llvm}, GCC~\cite{novillo2003tree}, but also research compilers such as Cetus~\cite{johnson2004experiences}.
While most SSA-based compilers today use the concept of imperative branching transitioning between sequences of instruction blocks, Click and Cooper~\cite{click1995combining} removed this restriction by introducing the sea-of-nodes concept of graph-based intermediate languages where all freely floating instructions are only constrained by explicit control and data dependences.
This concept is used in research compilers such as libfirm~\cite{braun2011firm} or Google's TurboFan JavaScript compiler.\footnote{\url{http://v8.dev}}
While the above approaches have shown the benefit of, especially SSA-based, intermediate representations the above concepts all aim for the generation of executable software, but do not target hardware design.
Graph-based \glspl{ir} certainly serve as inspiration to our \llhd[3]{}, but existing compilers use graph-based \glspl{ir} mostly for software compilation.
Nevertheless, there are first efforts to define intermediate representations for hardware designs, which we will discuss in detail in the following sections.

%-------------------------------------------------------------------------------
\subsection{FIRRTL}
\label{sec:relwork_firrtl}

FIRRTL~\cite{izraelevitz2017reusability} is the \gls{ir} most closely related to \llhd{} known to us.
FIRRTL acts as an abstraction layer between the Chisel~\cite{bachrach2012chisel} hardware generation framework and subsequent transformation passes and synthesis.
FIRRTL's semantics are closely coupled to those of Chisel and focuses mainly on the synthesis portion of the design flow.
Notable exceptions are support for certain testbench constructs (see \secref{sec:eval_irs}).
We identify the following four fundamental differences between \llhd{} and FIRRTL:

\textbf{(1)}
FIRRTL's fundamental data structure
is an \gls{ast}~\cite{li2016firrtl}.
Nodes may be assigned multiple times and \emph{at different granularities}, making identification of a value's single producer difficult.
In contrast, algorithms operating on \gls{ssa} forms prevail in modern compilers, and most research on transformations focuses on \gls{ssa}.
While low FIRRTL also requires \gls{ssa} form, algorithms \emph{requiring} this form are precluded from operating on all but the lowest level of abstraction.

\textbf{(2)}
FIRRTL cannot represent testbench and simulation constructs such as precise delays, events, and queues and dynamic arrays, as well as arbitrary programs that stimulate and check a circuit.
These constructs are essential to fully represent industry-standard \glspl{hdl} such as SystemVerilog and to transport designs written in such languages through the full digital design flow.

\textbf{(3)}
FIRRTL does not represent four- or nine-valued logic as defined by IEEE~1364 and IEEE~1164, respectively.
These types of logic model the additional states of a physical signal wire (beyond the fundamental 0 and 1) and captures concepts such as driving strength and impedance.
Industry-standard hardware designs in SystemVerilog and VHDL rely on this modeling capability to describe bidirectional signals, propagate unknown values, identify driving conflicts, and describe optimization opportunities during logic synthesis.

\textbf{(4)}
FIRRTL has the concept of ``three forms'' but does not clearly define their boundaries and to which parts of the design flow they apply~\cite{izraelevitz2017reusability}.
It merely states that low FIRRTL maps directly to Verilog constructs with straightforward semantics.

Overall we observe that \llhd{} is a \emph{superset} of FIRRTL.
Since \llhd{} provides a Turing-complete modeling construct for a circuit, and FIRRTL does not, any FIRRTL representation can be translated into an equivalent \llhd{} representation, but not vice-versa.
And by the same reasoning, modern \glspl{hdl} such as SystemVerilog and VHDL cannot be fully mapped to FIRRTL.

%-------------------------------------------------------------------------------
\subsection{Verification IRs}
\label{sec:relwork_verif}

CoreIR~\cite{mattarei2018cosa} focuses on verification.
It is designed to interact with higher-level functional descriptions of a circuit, such as Halide~\cite{ragan2013halide} or Verilog (via Yosys~\cite{wolf2018rtlil}), and represent these within a formal verification infrastructure.
Additional steps allow designs to be mapped to \glspl{cgra}.
CoreIR is geared specifically towards verification and deployment on \gls{fpga}-like devices, and as such does not cater to a full chip design flow.

%-------------------------------------------------------------------------------
\subsection{Synthesis IRs}
\label{sec:relwork_synth}

Many \glspl{ir} beside FIRRTL are engineered to interact with hardware synthesizers.
LNAST~\cite{wang2019lnast} targets the representation of the synthesizable parts of an \gls{hdl} circuit description in a language-agnostic \gls{ast}.
RTLIL~\cite{wolf2018rtlil} is part of the ``Yosys'' open-source tool suite and focuses mainly on logic synthesis.
It cannot represent higher-level constructs such as aggregate types or conditional assignment.
µIR~\cite{sharifian2019muir} is geared towards \gls{hls} design flows and tries to capture entire accelerator architectures.
These \glspl{ir} are very specifically crafted to transport designs into a synthesizer and focus solely on this part of the flow.

%-------------------------------------------------------------------------------
\subsection{Netlist IRs}
\label{sec:relwork_netlist}

\glspl{ir} exist that aim at capturing the gate-level netlist of a circuit.
LGraph~\cite{wang2019lgraph} is an open-source graph representation of such a circuit, together with additional aspects of the physical design flow such as cell libraries, timing and power characteristics, and placement information.
NetlistDB~\cite{anon2019netlistdb} follows a similar goal.
These \glspl{ir} cater only to the very end of the hardware design flow.

\section{Conclusion}
\label{sec:tail}

We showed with \moore{} and \llhd{} that a small and concise multi-level intermediate representation can represent the complex semantics of real-world VHDL and SystemVerilog circuits throughout the complete hardware design process.
Thanks to a novel three-level \gls{ir}, we present a hardware design language that is effective from simulation, over formal verification, to synthesized netlists.
We demonstrate the effectiveness of our design by outperforming commercial simulators on a variety of simulation tasks.
We expect that our concise and well-defined hardware design language, crafted in the spirit of the most successful \gls{ssa}-based compiler \glspl{ir}, both minimal and still expressive enough for real-world use cases, will provide the foundation for an open hardware design stack that follows the impressive evolution of modern compilers in recent years.

%%%%%%%%%%%%%%%%%%%%%%
%%   Bibliography   %%
%%%%%%%%%%%%%%%%%%%%%%

\bibliographystyle{ACM-Reference-Format}
\balance
\bibliography{bibliography}

\end{document}
\endinput